\documentclass[lettersize,journal]{IEEEtran}
\usepackage{amsmath,amsfonts}
\usepackage{algorithmic}
\usepackage{array}
\usepackage[caption=false,font=normalsize,labelfont=sf,textfont=sf]{subfig}
\usepackage{textcomp}
\usepackage{stfloats}
\usepackage{url}
\usepackage{verbatim}
\usepackage{graphicx}
\def\BibTeX{{\rm B\kern-.05em{\sc i\kern-.025em b}\kern-.08em
		T\kern-.1667em\lower.7ex\hbox{E}\kern-.125emX}}
\usepackage{balance}
\begin{document}
	\newtheorem{theorem}{Theorem}
\newtheorem{definition}{Definition}
\newtheorem{lemma}{Lemma}
\newtheorem{example}{Example}
\newtheorem{proposition}{Proposition}
\newtheorem{corollary}{Corollary}
\newtheorem{remark}{Remark}
\title{The non-GRS properties for the twisted generalized Reed-Solomon code and its extended code}

\author{ Canze Zhu,~~ Qunying Liao
\thanks{Canze Zhu is with the College of Mathematical Science, Sichuan Normal University, Chengdu Sichuan, 610066 (canzezhu@163.com).}
\thanks{Qunying Liao is with the College of Mathematical Science, Sichuan Normal University, Chengdu Sichuan, 610066 (qunyingliao@sicnu.edu.cn).}
\thanks{The second author is supported by National Natural Science Foundation of China (Grant No. 12071321).}}


\maketitle

\begin{abstract}In 2017, Beelen et al. firstly introduced  twisted generalized Reed-Solomon (in short, TGRS) codes, and constructed a large subclass of MDS TGRS codes.  Later, they proved that TGRS code is non-GRS  when the code rate  is less than one half. In this letter, basing on the dual code of the TGRS code or the extended TGRS code,  by using the Schur product, we prove that almost all of TGRS codes and extended TGRS codes are non-GRS when the code rate more than one half.
\end{abstract}


\section{Introduction}
\IEEEPARstart{T}{hroughout} this paper, let $\mathbb{F}_q$ be the finite field with $q$ elements, where $q$ is the power of  a prime. An $[n,k,d]$ linear code $\mathcal{C}$ over $\mathbb{F}_q$ is a $k$-dimensional subspace of $\mathbb{F}_q^n$ with minimum (Hamming) distance $d$ and length $n$. If the parameters reach the Singleton bound, namely, $d=n-k+1$, then $\mathcal{C}$  is  maximum distance separable (in short, MDS). Especially, generalized Reed-Solomon (in short, GRS) codes are a well known class of MDS codes, it is very important in coding theory and applications \cite{HZ21,10,19,25,40,LCD1,LCD2, LCD3,R0,R1,R2,R3,R4,R5}.  Other known MDS codes have been constructed from $n$-arcs in projective geometry \cite{MS77}, circulant matrices \cite{32}, Hankel matrices \cite{32}, or twisted Reed-Solomon (in short, TGRS) codes \cite{2,2022}.  

In 2017,  inspired by the construction for twisted Gabidulin codes \cite{t26}, Beelen et al. firstly introduced TGRS codes, which is a generalization for GRS codes, they also showed that TGRS codes could be well decoded.  Different from GRS codes, they showed that a  TGRS code is not necessarily MDS and presented a  sufficient and necessary condition for a TGRS code to be MDS  \cite{2,3,2022}. Especially, the authors showed that TGRS codes are not GRS when the code rate is less than one half \cite{2022}. Later, by TGRS codes, Lavauzelle et al. presented an efficient key-recovery attack used in the McEliece cryptosystem \cite{24}. TGRS codes are also used to construct LCD MDS codes by their applications in cryptography \cite{16,26}. Recently, the authors gave the parity check matrix for the TGRS code and obtained some self-dual TGRS codes with small Singleton defect \cite{t0,ZC21}. More relative results about self-orthogonal TGRS codes can be seen in \cite{SD0,SD1,SD2}.

In this letter, we focus on the non-GRS properties for the TGRS code and its extended code. Let $\mathcal{C}$ be  the TGRS code or extended TGRS (in short, ETGRS) code with code rate more than one half,  by calculating the dimension and the minimun Hamming distance of the Schur square for $\mathcal{C}^{\perp}$ under some assumptions, we  show that  $\mathcal{C}^{\perp}$ is non-GRS, and then $\mathcal{C}$ is non-GRS.  The rest of this letter is organized as follows. In section 2, some basic notations and results about linear codes are given. In section 3, we show that almost all of the TGRS code and ETGRS code are non-GRS. In section 4,  we conclude the whole paper.
\section{Preliminaries}
In this section, we  review some basic  knowledge.
\subsection{The dual codes, the Schur product and  equivalence for linear codes }
The nonation of the dual code is given in the following.

For $\mathbf{a}=(a_1,\ldots,a_n)$, $\mathbf{b}=(b_1,\ldots,b_n)$ $\in\mathbb{F}_q^n$, the inner product is defined as $\langle \mathbf{a},\mathbf{b}\rangle=\sum\limits_{i=1}^{n}a_ib_i.$ And then 
the dual code of $\mathcal{C}$ is defined as $$\mathcal{C}^{\perp}=\{\mathbf{c}^{'}\in\mathbb{F}_q^n~|~\langle\mathbf{c}^{'},\mathbf{c}\rangle=0~~ \text{for any}~\mathbf{c}\in\mathcal{C}\}.$$

The Schur product  is defined as follows.
\begin{definition}[\cite{R5}]\label{SP}
	For $\mathbf{x}=(x_1,\ldots,x_n ), \mathbf{y}= (y_1 ,\dots, y_n)\in \mathbb{F}_q^{n}$,  the Schur product  between $\mathbf{x}$ and $\mathbf{y}$ is
	defined as $\mathbf{x}\star\mathbf{y}:=(x_1y_1,\ldots,x_ny_n).$  The Schur product  of two $q$-ary codes $\mathcal{C}_1$ and $\mathcal{C}_2$ with length $n$ is
	defined as
	\begin{align*}
	\mathcal{C}_1\star\mathcal{C}_2=\langle \mathbf{c}_1 \star\mathbf{c}_2~|~\mathbf{c}_1\in\mathcal{C}_1,\mathbf{c}_2\in\mathcal{C}_2\rangle.
	\end{align*}
	Especially, for a code $\mathcal{C}$, we call $\mathcal{C}^2:=\mathcal{C}\star\mathcal{C}$ the Schur square of $\mathcal{C}$.
\end{definition} 
\begin{remark}\label{r1}For any linear code	$\mathcal{C}=\langle \boldsymbol{v}_1,\ldots,\boldsymbol{v}_{k}\rangle$ with $\boldsymbol{v}_i\in\mathbb{F}_q^{n}~(i=1,\ldots,k)$, we have
	\begin{align}\label{S1}
	\mathcal{C}^2=\langle \boldsymbol{v}_i\star\boldsymbol{v}_j~(i,j\in\{1,\ldots,k\})\rangle.
	\end{align}
\end{remark}

The definition of the equivalence for linear codes is given in the following.
\begin{definition} ({\cite{17}})\label{equi}
	Let $\mathcal{C}_1$ and $\mathcal{C}_2$ be $q$-ary linear codes with length $n$. We say that $\mathcal{C}_1$ and $\mathcal{C}_2$ are equivalent if
	there is a  permutation $\pi$ in the permutation group with order $n$ and $\boldsymbol{v}=(v_1,\dots,v_n)\in(\mathbb{F}_q^*)^{n}$ such that $\mathcal{C}_2=\Phi_{\pi,\boldsymbol{v}}(\mathcal{C}_1)$, where
	\begin{align*}
	\Phi_{\pi,\boldsymbol{v}}:\mathbb{F}_q^n\to\mathbb{F}_q^n,\quad(c_1,\ldots,c_n)\mapsto (v_1c_{\pi(1)},\ldots,v_nc_{\pi(n)}).\\
	\end{align*}
\end{definition}

\begin{remark}\label{es}By Definition \ref{equi}, if $\mathcal{C}_1$ and $\mathcal{C}_2$ are equivalent, then 
	
	$\bullet$  $\mathcal{C}_1^2$ and $\mathcal{C}_2^2$ are equivalent;
	
	$\bullet$ $\mathcal{C}_1^{\perp}$ and $\mathcal{C}_2^{\perp}$ are equivalent.
\end{remark}
\subsection{The GRS, TGRS and ETGRS code}
The definition of the GRS code is given in the following.
\begin{definition}[\cite{17}]\label{d0} 
	Let $\boldsymbol{\alpha}=(\alpha_1,\ldots,\alpha_n)\in\mathbb{F}_q^n$ with $\alpha_i\neq \alpha_j$ $(i\neq j)$ and
	$\boldsymbol{v} = (v_1,\ldots,v_{n})\in (\mathbb{F}_q^{*})^{n}$. Then
	the GRS code is defined as
	{\small	\begin{align*}
		&\mathcal{GRS}_{k,n}(\boldsymbol{\alpha},\boldsymbol{v})\\
		=&\{(v_1f({\alpha_{1}}),\ldots,v_nf(\alpha_{n}))|f(x)\in\mathbb{F}_q[x],\deg f(x)\le k-1\}.
		\end{align*}	}
\end{definition}

The dual code of the GRS code is given in the following.
\begin{lemma}[\cite{19}]\label{drs}
	Let $\boldsymbol{u}=(u_1,\ldots,u_n)$ with $u_j=\prod\limits_{i=1,i\neq j}^{n}(\alpha_j-\alpha_i)^{-1}$ $(j=1,\ldots,n)$, then
	$$\big(\mathcal{GRS}_{k,n}(\boldsymbol{\alpha},\boldsymbol{1})\big)^{\perp}=\mathcal{GRS}_{n-k,n}(\boldsymbol{\alpha},\boldsymbol{u}).$$
\end{lemma}

By the definition of the GRS code, Lemma \ref{drs} and Remark \ref{r1}, we have 
\begin{proposition}\label{pr}For $\frac{n}{2}\le k<n$, let $\boldsymbol{u}=(u_1,\ldots,u_n)$ with {\small $u_j=-\prod\limits_{i=1,i\neq j}^{n}(\alpha_j-\alpha_i)~(j=1,\ldots,n),$} then $
	\big(\mathcal{GRS}_{k,n}^{\perp}(\boldsymbol{\alpha},\boldsymbol{1})\big)^2=\mathcal{GRS}_{2(n-k)-2,n}(\boldsymbol{\alpha},\boldsymbol{u}^2)$ with dimension $2(n-k)-1$ and minimun Hamming distance $2k-n+2$.\\ 
\end{proposition}
The definition of the twisted polynomials linear  space 	$\mathcal{V}_{k,t,h,\eta}$ is given in the following.
\begin{definition}[\cite{2}]\label{d23} 
	Let $\eta\in\mathbb{F}_q^{*}$, and $t$, $h$, $k$ $\in\mathbb{N}$ with $0\le h<k\le q$. Then the set of $(k,t,h,\eta)$-twisted polynomials is defined as 
	{	\begin{align*}
		&\mathcal{V}_{k,t,h,\eta}\\
		=&\big\{f(x)=\sum\limits_{i=0}^{k-1}a_ix^i+\eta a_hx^{k-1+t}|a_i\in\mathbb{F}_q~(i=0,\ldots,k-1)\big\},
		\end{align*}}
	which is a $k$-dimensional $\mathbb{F}_q$-linear subspace. $h$ and $t$ are the hook and the twist, respectively.\\
\end{definition}

From the twisted polynomials linear  space 	$\mathcal{V}_{k,t,h,\eta}$, the definitions of the TGRS code and the ETGRS code are given in the following, respectively.
\begin{definition}[\cite{2}]\label{d1} 
	For any  $t$, $h$, $k$, $n$ $\in \mathbb{N}$ with $0\le h\le k-1<k-1+t<n\le q$, let $\eta\in\mathbb{F}_q^{*}$, $\boldsymbol{\alpha}=(\alpha_1,\ldots,\alpha_n)\!\in\!\mathbb{F}_q^n$ with $\alpha_i\neq \alpha_j$ $(i\neq j)$ and
	$\boldsymbol{v} = (v_1,\ldots,v_{n})\in (\mathbb{F}_q^{*})^{n}$. Then
	the  TGRS code is defined as
	\begin{align*}
	\mathcal{C}_{t,h,k,n}(\boldsymbol{\alpha},\boldsymbol{v},\eta)
	=\{(v_1f({\alpha_{1}}),\ldots,v_nf(\alpha_{n}))~|~f(x)\in \mathcal{V}_{k,t,h,\eta}\}.
	\end{align*}	
	For $h>0$, the  ETGRS code is defined as
	\begin{align*}
	&\mathcal{C}_{t,h,k,n}(\boldsymbol{\alpha},\boldsymbol{v},\eta,\infty)\\
	=&\{(v_1f({\alpha_{1}}),\ldots,v_nf(\alpha_{n}),f_{h})~|~f(x)\in  \mathcal{V}_{k,t,h,\eta}\},
	\end{align*}	
	where  $f_{h}$ is the coefficient of $x^{h}$ in $f(x)$. Especially, $\mathcal{C}_{t,h,k,n}(\boldsymbol{\alpha},\boldsymbol{1},\eta)$ and $\mathcal{C}_{t,h,k,n}(\boldsymbol{\alpha},\boldsymbol{1},\eta,\infty)$ are called  the  TRS code and the  ETRS code, respectively.\\
\end{definition}

\begin{remark}
	$\bullet$ Note that $f(0)=f_0$, thus it is not necessary to define the  ETGRS code when hook $h=0$. 
	
	$\bullet$ In \cite{2},  $\mathcal{C}_{t,h,k,n}(\boldsymbol{\alpha},\boldsymbol{v},\eta)$ and $\mathcal{C}_{t,h,k,n}(\boldsymbol{\alpha},\boldsymbol{v},\eta,\infty)$ are called TGRS codes without difference. But it is not different from Definition \ref{d1}.\\
\end{remark}
Beelen et al. gave the non-GRS property for the TGRS code as the following.
\begin{lemma}[Theorem $30$~\cite{2022}]\label{2k}
	If $3<k<\frac{n}{2}$, then	$\mathcal{C}_{t,h,k,n}(\boldsymbol{\alpha},\boldsymbol{v},\eta)$ is non-GRS.
\end{lemma}

Note that any punctured code of the GRS code is a GRS code and $\mathcal{C}_{t,h,k,n}(\boldsymbol{\alpha},\boldsymbol{v},\eta)$ is a punctured code of $\mathcal{C}_{t,h,k,n}(\boldsymbol{\alpha},\boldsymbol{v},\eta,\infty)$. By Lemma \ref{2k}, we have 
\begin{lemma}\label{2k1}
	If $3<k<\frac{n+1}{2}$, then {\small$\mathcal{C}_{t,h,k,n}(\boldsymbol{\alpha},\boldsymbol{v},\eta,\infty)$} is non-GRS.
\end{lemma}

The authors gave the parity check matrix for $\mathcal{C}_{t,h,k,n}(\boldsymbol{\alpha},\boldsymbol{v},\eta)$, which is useful to show that  $\mathcal{C}_{t,h,k,n}(\boldsymbol{\alpha},\boldsymbol{v},\eta)$ is non-GRS when $2k\ge n$.
{
	\begin{lemma}[Theorem $3.1$ \cite{ZC21}]\label{dual0}
		For any $m\in\mathbb{N}$, let {\small $$u_j=\prod_{i=1,i\neq j}^{n}(\alpha_j-\alpha_{i})^{-1}~(i=1,\ldots,n),~~L_m=\sum\limits_{l=1}^{n}u_l\alpha_{l}^{n-1+m}$$
			$$\text{and}~~~\tilde{L}=\sum\limits_{m=1}^{k-h-1}L_{m} L_{k+t-h-1-m}-\eta^{-1}(1+\eta L_{k+t-h-1}).$$} Then $\mathcal{C}_{t,h,k,n}(\boldsymbol{\alpha},\boldsymbol{v},\eta,\infty)$ has the parity check matrix  $$\tilde{H}_{n-k}(\boldsymbol{v})=\left(\begin{matrix}\tilde{\boldsymbol{\beta}}_1,\tilde{\boldsymbol{\beta}}_2,\ldots,\tilde{\boldsymbol{\beta}}_n
		\end{matrix}\right),$$ 
		where {\footnotesize
			\begin{align*}
			\tilde{\boldsymbol{\beta}}_{j}
			=\left(\begin{matrix}
			&\!\!\!\!\frac{u_j}{v_j}\\
			&\!\!\!\!\vdots\\
			&\!\!\!\!\frac{u_j}{v_j}\alpha_j^{n-(k+t+1)}\\
			&\!\!\!\!\frac{u_j}{v_j}\big(\tilde{L}\alpha_j^{n-k-t}+\alpha_j^{n-(h+1)}-\sum\limits_{m=1}^{k-h-1}L_{m}\alpha_j^{n-(h+1)-m}\big)\\
			&\!\!\!\!\frac{u_j}{v_j}\big(\alpha_j^{n-(k+t-1)}-L_1\alpha_j^{n-k-t}\big)\\
			&\!\!\!\!\vdots\\
			&\!\!\!\!\frac{u_j}{v_j}\big(\alpha_j^{n-(k+1)}- L_{t-1}\alpha_j^{n-k-t}\big)\\
			\end{matrix}\right)_{(n-k)\times n}.
			\end{align*}}\\
\end{lemma}}

{
	\begin{remark}\label{duall}
		Under the assumption in Lemma $\ref{dual0}$, let
		\begin{align*}
		&\mathcal{V}_{k,t,h,\eta}^{\perp}\\
		=&\Big\{f(x)=\sum\limits_{i=0}^{n-(k+t+1)}a_ix^i+\sum_{j=0}^{t-1}a_{n-k-t+j}h_{n-k-t+j}(x)~\Big|\\
		&\qquad\quad a_i\in\mathbb{F}_q~(i=0,\ldots,n-k-1)\Big\}
		\end{align*}
		with{\footnotesize \begin{align*}
			&h_{n-k-t+j}(x)\\
			=&\begin{cases}
			x^{n-(h+1)}+\tilde{L}x^{n-k-t}-\sum\limits_{m=1}^{k-h-1}L_{m}x^{n-(h+1)-m},&\text{if~}j=0;\\
			x^{n-k-t+j}-L_jx^{n-k-t},&\!\!\!\!\!\!\!\!\!\!\text{if~}~1\le j\le t-1.
			\end{cases}
			\end{align*}}
		Then{\small$$\mathcal{C}_{t,h,k,n}^{\perp}(\boldsymbol{\alpha},\boldsymbol{1},\eta)=\Big\{\big(u_1g({\alpha_{1}}),\ldots,{u_n}g(\alpha_{n})\big)|g(x)\in \mathcal{V}^{\perp}_{k,t,h,\eta}\Big\}.$$}
\end{remark}}
\section{Non-GRS properties for the TGRS code and ETGRS code } 
In this section, by using the Schur product, we prove that almost all of TGRS codes and ETGRS codes are non-GRS when the code rate more than one half.

\begin{theorem}\label{tt}Let  $t$, $h$, $k$, $n$ $\in \mathbb{N}$ with $0\le h\le k-1$ and $k+t\le n\le q$. If $\frac{n}{2}\le k\le n-3$ and one of the following conditions $(1.1)$-$(1.6)$ holds, then $\mathcal{C}_{t,h,k,n}(\boldsymbol{\alpha},\boldsymbol{v},\eta)$ is  non-GRS.
	
	$(1.1)$ $n=k+t$, $2k\ge n+2$, and $3\le h\le k- 3$;
	
	$(1.2)$ $n=k+t+1$, $2k\ge n+2$, and $2\le h\le k- 3$;
	
	$(1.3)$ $n\ge k+t+2$, $2k\ge n$, $t=1$ and $2\le h\le k-2$;
	
	$(1.4)$ $n\ge k+t+2$, $2k\ge n+1$, $t=2$ and $1\le h\le k-3$;
	
	$(1.5)$ $n\ge k+t+2$, $2k\ge n$, $t\ge 3$ and $h\in\{0,\ldots,k-1\}\backslash\{k-t\}$;
	
	$(1.6)$ $n\ge k+t+2$, $2k\ge n+1$, $t\ge 3$ and $h=k-t$.\\
\end{theorem}

{\bf Proof}. 
We show that $\dim\big(\big(\mathcal{C}^{\perp}_{t,h,k,n}(\boldsymbol{\alpha},\boldsymbol{1},\eta)\big)^2\big)\ge 2(n-k)$ as follows.

By Remark \ref{duall}, we have
{
	\begin{align*}
	\mathcal{C}_{t,h,k,n}^{\perp}(\boldsymbol{\alpha},\boldsymbol{1},\eta)\!\!
	=\big\{\big(u_1f({\alpha_{1}}),\ldots,u_nf(\alpha_{n})\big)\big|f(x)\in \mathcal{V}^{\perp}_{k,t,h,\eta}\big\},\!\!
	\end{align*}
	where{\small \begin{align*}
		\mathcal{V}_{k,t,h,\eta}^{\perp}
		=&\Big\{f(x)=\sum\limits_{i=0}^{n-(k+t+1)}a_ix^i+\sum_{j=0}^{t-1}a_{n-k-t+j}h_{n-k-t+j}(x)|\\
		&\qquad\qquad a_i\in\mathbb{F}_q~(i=0,\ldots,n-k-1)\Big\}
		\end{align*}}
	with {\small \begin{align}\label{hhh}\begin{aligned}
		&h_{n-k-t+i}(x)\\
		=&\begin{cases}
		x^{n-(h+1)}+\tilde{L}x^{n-k-t}-\sum\limits_{m=1}^{k-h-1}L_{m}x^{n-(h+1)-m},&\!\!\!\!\!\!\!\text{~if~}i=0;\\
		x^{n-k-t+i}-L_ix^{n-k-t},&\!\!\!\!\!\!\!\!\!\!\!\!\!\!\!\!\!\!\!\!\!\!\!\!\!\!\!\!\!\!\!\text{~if~}1\le i\le t-1.
		\end{cases}\end{aligned}
		\end{align} }
	Thus
	\begin{align*}
	\big(\mathcal{C}_{t,h,k,n}^{\perp}(\boldsymbol{\alpha},\boldsymbol{1},\eta)\big)^2\!\!
	=\!\!\Big\{\big(g({\alpha_{1}}),\ldots,g(\alpha_{n})\big)\big|g(x)\in(\mathcal{V}_{k,t,h,\eta}^{\perp})^{2}\Big\}
	\end{align*}
	with 
	\begin{align*}
	(\mathcal{V}_{k,t,h,\eta}^{\perp})^{2}
	=\big\langle g(x)=f_1(x)f_2(x)\big|~f_1(x),f_2(x)\in\mathcal{V}_{k,t,h,\eta}^{\perp}\big\rangle.
	\end{align*}
	
	Now we show that if one of the conditions $(1.1)$-$(1.6)$ holds, then there exists $g_i(x) \in(\mathcal{V}_{k,t,h,\eta}^{\perp})^{2}$ $(i=0,\ldots,2(n-k)-1)$ such that
	\begin{align}\label{gi}
	\deg g_i(x)\le n-1~\text{and}~\deg g_i(x)\neq \deg g_{j}(x)~(i\neq j).
	\end{align}
	
	$(1.1)$ For $n=k+t$, $2k\ge n+2$, and $3\le h\le k- 3$, we have
	{\small \begin{align*}
		&(\mathcal{V}_{k,t,h,\eta}^{\perp})^{2}\\
		=&\langle h_{l}(x)h_{s}(x),~h_{0}(x)h_{i}(x),~h_{0}^2(x)~(l,s,i\in\{1,\ldots,n-k-1\})\rangle.
		\end{align*} }By (\ref{hhh}), one has
	\begin{align}\label{h1}
	\begin{aligned}
	&\deg\big(  h_{l}(x)h_{s}(x)\big) = l+s\le 2(n-k)-2\le n-1~\\
	&(l,s\in \{1,\ldots,n-k-1\}).
	\end{aligned}	
	\end{align}
	Furthermore, for $h_{0}(x)h_{i}(x)$, we have the following two cases.\\
	If $3\le h \le 2k-n+1$, then $2(n-k)-2\le n-(h+1)\le n-4 $. Thus
	\begin{align}\label{h11}\begin{aligned}
	2(n-k)-1&\le \deg\big( h_{0}(x)h_{i}(x)\big)\\
	&=n-(h+1)+i\\
	&\le n-1~(i\in\{1,2,3\}).
	\end{aligned}
	\end{align}
	If $2k-n+2 \le h\le k-3$, then 
	$1\le n-2k+h<n-2k+h+1<n-2k+h+2\le n-k-1.$ Now by $2k\ge n+2$, one has \begin{align}\label{h12}\begin{aligned}
	2(n-k)-1\le& \deg\big(   h_{0}(x)h_{n-2k+h-1+i}(x)\big)\\
	=&2(n-k)-2+i\le n-1~(i\in\{1,2,3\}).\end{aligned}
	\end{align}
	So far, by $(\ref{h1})$-$(\ref{h12})$, $(\ref{gi})$ holds.\\

	$(1.2)$ For $n=k+t+1$, $2k\ge n+2$, and $2\le h\le k- 3$, we have\begin{align*}
	(\mathcal{V}_{k,t,h,\eta}^{\perp})^{2}=&\langle 1, h_0(x), h_m(x), h_{l}(x)h_{s}(x),h_{0}(x)h_{i}(x),~h_{0}^2(x)\\
	&\qquad\qquad\qquad(m,l,s,i\in\{2,\ldots,n-k-1\})\rangle.
	\end{align*}
	By (\ref{hhh}), one has
	{\small \begin{align}\label{h2}
		\begin{aligned}
		&\deg h_{m}(x)= m~(m=2,3),\\
		&4\le \deg\big( h_{l}(x)h_{s}(x)\big) = l+s\le n-1~(l,s\in \{2,\ldots,n-k-1\}).
		\end{aligned}
		\end{align}}
	Furthermore, for $h_{0}(x)h_{i}(x)$, we have the following two cases.\\
	If $3\le h \le 2k-n+1$, then $2(n-k)-2\le n-(h+1)\le n-4 $. Thus
	\begin{align}\label{h21}	\begin{aligned}
	2(n-k)-1&\le \deg\big( h_{0}(x)h_{i}(x)\big)\\
	&=n-(h+1)+i\le n-1~(i\in\{2,3\}).	\end{aligned}
	\end{align}
	If $2k-n+2 \le h\le k-3$, then $2\le n-2k+h+1<n-2k+h+2\le n-k-1.$ Thus 
	\begin{align}\label{h22}\begin{aligned}
	2(n-k)&\le \deg\big(h_{0}(x)h_{n-2k+h+i}(x)\big)\\
	&=2(n-k)-1+i\le n-1~(i\in\{1,2\}).	\end{aligned}
	\end{align}
	Now by $1\in(\mathcal{V}_{k,t,h,\eta}^{\perp})^{2}$ and $(\ref{h2})$-$(\ref{h22})$, $(\ref{gi})$ holds.\\
	
	$(1.3)$ For $n\ge k+t+2$, $2k\ge n$, $t=1$ and $2\le h\le k-2$, we have 
	\begin{align*}
	&(\mathcal{V}_{k,t,h,\eta}^{\perp})^{2}\\
	=&\langle x^{l+s},~x^{i}h_{0}(x),~h_{0}^2(x)~~~(l,s,i\in\{0,1,\ldots,n-k-2\})\rangle.
	\end{align*}  
	Obviously,
	\begin{align}\label{h3}
	\deg\big( x^{l+s}\big)= l+s\le n-1~(l,s\in \{0,1,\ldots,n-k-2\}).
	\end{align}
	Furthermore, for $x^jh_{i}(x)$, we have the following two cases.\\
	If $2\le h \le 2k-n+2$, then $2(n-k)-3\le n-(h+1)\le n-3$. Thus
	\begin{align}\label{h31}\begin{aligned}
	2(n-k)-3&\le \deg\big(x^{i} h_{0}(x)\big)\\
	&=n-(h+1)+i\le n-1~(i\in\{0,1,2\}).\end{aligned}
	\end{align}
	If $2k-n+3 \le h\le k-2,$  then 
	$h+1\le n-2k+h-2<n-2k+h-1<n-2k+h\le n-k-2,$ thus 
	\begin{align}\label{h32} \begin{aligned}
	2(n-k)-4\le &\deg\big(x^{n-2k+h-2+i}h_{0}(x)\big)\\
	=&2(n-k)-3+i\le n-1~(i\in\{0,1,2\}).\end{aligned}
	\end{align}
	Now by  $(\ref{h3})$-$(\ref{h32})$, $(\ref{gi})$ holds.\\
	
	$(1.4)$ For $n\ge k+t+2$, $2k\ge n+1$, $t=2$ and $1\le h\le k-3$, we have 
	{\small\begin{align*}
		(\mathcal{V}_{k,t,h,\eta}^{\perp})^{2}
		=&\big\langle x^{l+s},~x^{i}h_{0}(x),~x^{i}h_{1}(x),~h_{0}^2(x),~h_{0}(x)h_1(x),~h_1^2(x)\\
		&~~~(l,s,i\in\{0,1,\ldots,n-k-3\})\big\rangle.\end{align*}} By (\ref{hhh}), one has
{\small	\begin{align}\label{h4}\begin{aligned}
	&\deg\big( x^{l+s}\big)= l+s\le 2n-2k-6\le n-1\\
	&~~(l,s\in \{0,1,\ldots,n-k-3\}),\\
	&\deg(x^{n-k-5+m}h_{1}(x))=2(n-k)-6+m~(m=1,2),\\
	&\deg\big( h_{1}^2(x)\big)=2(n-k)-2. 
	\end{aligned}
	\end{align}}
	Furthermore, for $x^{l}h_{0}(x)$, we have the following two cases.\\	
	If $1\le h \le 2k-n$, then $2(n-k)-1\le n-(h+1)\le n-2$, thus
	\begin{align}\label{h41}\begin{aligned}
	2(n-k)-1&\le\deg\big(x^{i}h_{0}(x)\big)\\
	&=n-(h+1)+i\le n-1~(i\in\{0,1\}).\end{aligned}
	\end{align}
	If $2k-n+1\le h\le k-3,$  then $h-1\le n-2k+h-2<n-2k+h\le n-k-3,$
	thus{\small \begin{align}\label{h42}\begin{aligned}
		&\deg\big(x^{n-2k+h-2}h_{0}(x)\big)=2(n-k)-3,\\
		& \deg\big(x^{n-2k+h}h_{0}(x)\big)=2(n-k)-1\le n-1.
		\end{aligned}
		\end{align}}
	Now by  $(\ref{h4})$-$(\ref{h42})$, $(\ref{gi})$ holds.\\
	
	$(1.5)$ For $n\ge k+t+2$, $2k\ge n$ and $h\in\{0,\ldots,k-1\}\backslash\{k-t\}$, we have 
	{\small\begin{align*}
		& (\mathcal{V}_{k,t,h,\eta}^{\perp})^{2}\\
		=&\Big\langle x^{s_1+s_2},~x^{s}h_{n-k-t+l}(x),~h_{n-k-t+l_1}(x)h_{n-k-t+l_2}(x)\\
		&\big(s,s_1,s_2\in\{0,\ldots,n-k-t-1\},l,l_1,l_2\in\{0,1,\ldots,t-1\}\big)\Big\rangle,
		\end{align*}}and{\small
		\begin{align}\label{h5}\begin{aligned}
		&\deg\big( x^{l+s}\big)= l+s\le n-1~(l,s\in \{0,1,\ldots,n-k-t-1\}),\\
		&\deg(x^{n-k-t-2}h_{n-k-t+l}(x))=2(n-k)-2t-2+l\\
		&~(l=1,2,\ldots,t-1),\\
		&\deg\big(h_{n-k-3}(x)h_{n-k-t+l_1}(x)\big)=2(n-k)-t-3+l_1\\
		&~(l_1=1,\ldots,t-1),~\\
		&\deg\big(h_{n-k-2}(x)h_{n-k-1}(x)\big)=2(n-k)-3,\\
		&\deg\big(h_{n-k-1}^{2}(x)\big)=2(n-k)-2.\end{aligned}
		\end{align}}
	Furthermore, for $x^{l}h_{0}(x)$ and $h_{i}(x)h_0(x)$, we have the following three cases.\\	
	If $0\le h \le 2k-n$, then $2(n-k)-1\le n-(h+1)\le n-1$, thus
	\begin{align}\label{h51}
	\deg\big(h_{0}(x)\big)=n-(h+1)\le n-1.
	\end{align}
	If $2k-n+1\le h\le k-t-1,$  then  $h+1\le n-2k+h\le n-k-t-1,$
	thus \begin{align}\label{h52}
	\deg\big(x^{n-2k+h}h_{0}(x)\big)=2(n-k)-1\le n-1.
	\end{align}
	If $k-t+1\le h\le k-1 ,$  then $n-k-t+1\le n-2k+h\le n-k-1,$
	thus \begin{align}\label{h53}
	\deg\big(h_{n-2k+h}(x)h_{0}(x)\big)=2(n-k)-1\le n-1.
	\end{align}
	Now by  $(\ref{h5})$-$(\ref{h53})$, $(\ref{gi})$ holds.\\
	
	$(1.6)$ For $n\ge k+t+2$, $2k\ge n+1$ and $h=k-t$, in the similar proof as that for $(1.5)$,  we can get $(\ref{gi})$.\\
	
	Let $\mathcal{V}=\Big\{\sum\limits_{i=0}^{2(n-k)-1}a_ig_i(x)~|~a_i\in\mathbb{F}_q\Big\},$ where $g_i(x)$ $(i=0,\ldots,2(n-k)-1)$ is given in $(\ref{gi})$, then 
	$\dim (\mathcal{V})=2(n-k)$ and $\mathcal{V}\subseteq(\mathcal{V}_{k,t,h,\eta}^{\perp})^{2}.$
	Thus{\small $$\mathcal{C}_{\mathcal{V}}=\Big\{\big(g(\alpha_1),\ldots,g(\alpha_n)\big)~|~g(x)\in\mathcal{V}\Big\}\subseteq \big(\mathcal{C}_{t,h,k,n}^{\perp}(\boldsymbol{\alpha},\boldsymbol{1},\eta)\big)^2.$$}
	Now by $\deg (g(x))\le n-1$ $(~\forall g(x)\in\mathcal{V}~)$, we have $\big(g(\alpha_1),\ldots,g(\alpha_n)\big)\neq 0~~(\forall g(x)\in\mathcal{V}\backslash\{0\}).$
	It implies that $\dim\Big( \big(\mathcal{C}_{t,h,k,n}^{\perp}(\boldsymbol{\alpha},\boldsymbol{v},\eta)\big)^2\Big)\ge\dim(\mathcal{C}_{\mathcal{V}})=\dim{\mathcal{V}}=2(n-k).$
	
	By Propositon \ref{pr}, for any $[n,k]$  GRS code $\mathcal{C}$,   $\big(\mathcal{C}^{\perp}\big)^{2}=2(n-k)-1$. Thus if one of conditions $(1.1)$-$(1.6)$ holds, then $\mathcal{C}_{t,h,k,n}^{\perp}(\boldsymbol{\alpha},\boldsymbol{v},\eta)$ is non-GRS, and so $\mathcal{C}_{t,h,k,n}(\boldsymbol{\alpha},\boldsymbol{v},\eta)$ is non-GRS. $\hfill\Box$\\
}

The following lemma is necessary to determine some codewords in $\mathcal{C}_{t,h,k,n}^{\perp}(\boldsymbol{\alpha},\boldsymbol{v},\eta,\infty)$.
\begin{lemma}\label{lu}
	For any $m\in\mathbb{N}$ and $A\subseteq\mathbb{F}_q$ with $|A|>2$, let $L_A(m)=\sum\limits_{\alpha\in A}\alpha^{m}\prod_{\beta\in\mathbb{F}_q\backslash A}(\alpha-\beta),$ then
{\small	\begin{align}\label{lq}
	L_A(m)=\begin{cases}
	0,\quad&\text{if}~m\le |A|-2;\\
	-1,\quad&\text{if}~m=|A|-1;\\
	-\sum\limits_{\alpha\in A}\alpha,&\text{if}~m=|A|.
	\end{cases}
	\end{align}}
\end{lemma}

{\bf Proof.} For any $l\in\mathbb{Z}^{+}$, it is well-known that {\small
\begin{align}\label{lq1}
\sum\limits_{\beta\in \mathbb{F}_q}\beta^{l}=\begin{cases}
-1,\quad&\text{if}~q-1\mid l;\\
0,\quad&\text{otherwise}.
\end{cases}
\end{align}}
Note that{\small \begin{align*}
&\prod_{\beta\in\mathbb{F}_q\backslash A}(\alpha-\beta)\\
=&\alpha^{q-|A|}-\sum\limits_{\beta\in\mathbb{F}_q\backslash A}\beta\alpha^{q-|A|-1}+\cdots+(-1)^{q-|A|}\prod_{\beta\in\mathbb{F}_q\backslash A}\beta\\
=&\alpha^{q-|A|}+\sum\limits_{\gamma\in A}\gamma\alpha^{q-|A|-1}+\cdots+(-1)^{q-|A|}\prod_{\beta\in\mathbb{F}_q\backslash A}\beta,
\end{align*}}
thus we have {\small
	\begin{align}\label{lq2}	\begin{aligned}
	L_A(m)
	=&\sum\limits_{\alpha\in A}\alpha^{m}\prod_{\beta\in\mathbb{F}_q\backslash A}(\alpha-\beta)\\
	=&\sum\limits_{\alpha\in \mathbb{F}_{q}}\Big(\alpha^{q-|A|+m}+\sum\limits_{\gamma\in A}\gamma\alpha^{q-|A|-1+m}+\\
	&\qquad\qquad\qquad\cdots+(-1)^{q-|A|+m}\prod_{\beta\in\mathbb{F}_q\backslash A}\beta\alpha^m\Big).
	\end{aligned}\end{align}}
Now by $(\ref{lq1})$-$(\ref{lq2})$, we obtain $(\ref{lq})$. $\hfill\Box$\\
{
	\begin{theorem}\label{tte}
		For $t\ge 2$, $3\le k\le n-2$ and $n\ge k+t+1$, $\mathcal{C}_{t,h,k,n}(\boldsymbol{\alpha},\boldsymbol{v},\eta,\infty)$ is non-GRS.
\end{theorem}}

{\bf Proof}.  By Lemma \ref{2k1}, if $3\le k<\frac{n+1}{2}$, then $\mathcal{C}_{t,h,k,n}(\boldsymbol{\alpha},\boldsymbol{v},\eta,\infty)$ is non-GRS. Thus it is enought to prove that the theorem is true for $2k\ge n+1 $.

In the following, we prove that $\boldsymbol{c}_i\in\mathcal{C}_{t,h,k,n}^{\perp}(\boldsymbol{\alpha},\boldsymbol{v},\eta,\infty)$ $(i=1,2,3)$, where{\small
	\begin{align*}
	&\boldsymbol{c}_1=\Big(~\frac{u_1}{v_1}\alpha_1^{n-k-t-1}~,\ldots,\frac{u_n}{v_n}\alpha_{n}^{n-k-t-1},0~\Big),\\
	&\boldsymbol{c}_2=\Big(~\frac{u_1}{v_1}\alpha_1^{n-k-t}~,\ldots,\frac{u_n}{v_n}\alpha_{n}^{n-k-t},-\eta~\Big),\\
	&\boldsymbol{c}_3=\Big(~\frac{u_1}{v_1}\alpha_1^{n-k-t+1}~,\ldots,\frac{u_n}{v_n}\alpha_{n}^{n-k-t+1},-\eta \sum_{i=1}^{n}\alpha_i~\Big).
	\end{align*}}

Denote $A_{\boldsymbol{\alpha}}=\{\alpha_1,\ldots,\alpha_n\}$, for $s\in\{n-k-t-1,n-k-t,n-k-t+1\}$ and $l\in\{0,\ldots,k-1,k-1+t\}$, by the assumption $t\ge 2$ and Lemma \ref{lu}, we have
\begin{align*}
\begin{aligned}
&\sum\limits_{i=0}^{n}u_i\alpha_i^{s}\alpha_i^{l}\\
=&-\sum\limits_{\alpha\in A_{\boldsymbol{\alpha}}}\alpha^{s+l}\prod_{\beta\in\mathbb{F}_q\backslash A_{\boldsymbol{\alpha}}}(\alpha-\beta)\\
=&\begin{cases}
1, &\text{if}~s=n-k-t\text{~and~}l=k-1+t;\\
\sum_{i=1}^{n}\alpha_i,&\text{if}~s=n-k-t+1 \text{~and~}l=k-1+t; \\
0,&\text{otherwise}.
\end{cases} \end{aligned}
\end{align*}
By above equation, we can verify that $$\boldsymbol{c}_i\in\mathcal{C}_{t,h,k,n}^{\perp}(\boldsymbol{\alpha},\boldsymbol{v},\eta,\infty)~ (i=1,2,3)$$ directly. Thus
\begin{align*}
\boldsymbol{c}&=\boldsymbol{c}_1\star\boldsymbol{c}_3-\boldsymbol{c}_2\star\boldsymbol{c}_2\\
&=\big(0,0,\ldots,0,\eta^2~\big)\in (\mathcal{C}_{k,n}^{\perp}(\boldsymbol{\alpha},\boldsymbol{v},\eta,\infty))^{2}.
\end{align*}

By Proposition \ref{pr}, for an $[n+1,k]$ GRS code $\mathcal{C}$,   $(\mathcal{C}^{\perp})^2$ is with minimun Hamming distance
\begin{align*}
d=2k-(n+1)+2\ge 2.
\end{align*}
Thus  $\boldsymbol{c}\notin(\mathcal{C}^{\perp})^2$, and then $\mathcal{C}_{t,h,k,n}^{\perp}(\boldsymbol{\alpha},\boldsymbol{v},\eta,\infty)$ is non-GRS, it implies that  $\mathcal{C}_{t,h,k,n}(\boldsymbol{\alpha},\boldsymbol{v},\eta,\infty)$ is non-GRS.
$\hfill\Box$\\

\begin{remark} $\bullet$ Note that any punctured code of the GRS code is a GRS code and $\mathcal{C}_{t,h,k,n}(\boldsymbol{\alpha},\boldsymbol{v},\eta)$ is a punctured code of $\mathcal{C}_{t,h,k,n}(\boldsymbol{\alpha},\boldsymbol{v},\eta,\infty)$. Thus, if one of conditions $(1.1)$-$(1.6)$ in Theorem \ref{tt} holds, then $\mathcal{C}_{t,h,k,n}(\boldsymbol{\alpha},\boldsymbol{v},\eta,\infty)$ is non-GRS.
	
	$\bullet$ The assumption in Theorem \ref{tte} is weaker than that in Theorem \ref{tt} when $t\ge 2$ and $n\ge k+t+1$.
\end{remark}

\section{Conclusion}
In this letter, by using the Schur product, we prove that almost all of TGRS codes and ETGRS codes are non-GRS when the code rate more than one half.  However, if $n,k,t,h$ do not satisfy one of the conditions in Theorem \ref{tt}, we can not obtain the non-GRS property for the TGRS code by the method given in the proof for Theorem \ref{tt}. The main reason is that we can not obtain $\dim\big(\big(\mathcal{C}^{\perp}_{t,h,k,n}(\boldsymbol{\alpha},\boldsymbol{1},\eta)\big)^2\big)\ge 2k$  without the conditions in Theorem \ref{tt}.



\newpage

\end{document}